\documentclass[apjl]{emulateapj}
\usepackage{color}

\newcommand{\Msun}{M_{\odot}}

\def\gsim{\mathrel{\rlap{\lower 4pt \hbox{\hskip 1pt $\sim$}}\raise 1pt \hbox {$>$}}} \def\lsim{\mathrel{\rlap{\lower 4pt \hbox{\hskip 1pt $\sim$}}\raise 1pt \hbox {$<$}}}   

\shorttitle{The extremely luminous Type Ia SN 2009dc} \shortauthors{Yamanaka et al.}

\begin{document}

\title{Early phase observations of extremely luminous Type Ia Supernova 2009dc}

\author{M. \textsc{Yamanaka}\altaffilmark{1,2},
        K. S. \textsc{Kawabata}\altaffilmark{2},
        K. \textsc{Kinugasa}\altaffilmark{3},
        M. \textsc{Tanaka}\altaffilmark{4}, 
        A. \textsc{Imada}\altaffilmark{5},
        K. \textsc{Maeda}\altaffilmark{6}, 
        K. \textsc{Nomoto}\altaffilmark{6}, \\
        A. \textsc{Arai}\altaffilmark{1},
	S. \textsc{Chiyonobu}\altaffilmark{1},
        Y. \textsc{Fukazawa}\altaffilmark{1}, 
	O. \textsc{Hashimoto}\altaffilmark{3}, 
	S. \textsc{Honda}\altaffilmark{3}, 
        Y. \textsc{Ikejiri}\altaffilmark{1},
        R. \textsc{Itoh}\altaffilmark{1},
	Y. \textsc{Kamata}\altaffilmark{9}, \\
 	N. \textsc{Kawai}\altaffilmark{7}, 
        T. \textsc{Komatsu}\altaffilmark{1}, 
 	D. \textsc{Kuroda}\altaffilmark{5}, 
	H. \textsc{Miyamoto}\altaffilmark{1}
	S. \textsc{Miyazaki}\altaffilmark{9}
        O. \textsc{Nagae}\altaffilmark{1},
	H. \textsc{Nakaya}\altaffilmark{9},\\
        T. \textsc{Ohsugi}\altaffilmark{2},
	T. \textsc{Omodaka}\altaffilmark{8}, 
 	N. \textsc{Sakai}\altaffilmark{8}, 
        M. \textsc{Sasada}\altaffilmark{1}, 
	M. \textsc{Suzuki}\altaffilmark{10},
	H. \textsc{Taguchi}\altaffilmark{3},  
	H. \textsc{Takahashi}\altaffilmark{3},\\ 
        H. \textsc{Tanaka}\altaffilmark{1},
	M. \textsc{Uemura}\altaffilmark{2}, 
        T. \textsc{Yamashita}\altaffilmark{9},
	K. \textsc{Yanagisawa}\altaffilmark{5}, and
	M. \textsc{Yoshida}\altaffilmark{5}
}

\altaffiltext{1}{Department of Physical Science, Hiroshima University, Kagamiyama 1-3-1, Higashi-Hiroshima 739-8526, Japan; myamanaka@hiroshima-u.ac.jp} 
\altaffiltext{2}{Hiroshima Astrophysical Science Center, Hiroshima University, Higashi-Hiroshima, Hiroshima 739-8526, Japan} 
\altaffiltext{3}{Gunma Astronomical Observatory, Takayama, 
Gunma 377-0702, Japan} 
\altaffiltext{4}{Department of Astronomy, School of Science, University 
 of Tokyo, Bunkyo-ku, Tokyo 113-0033, Japan} 
\altaffiltext{5}{Okayama Astrophysical Observatory, National Astronomical 
Observatory of Japan, Kamogata, Asakuchi-shi, Okayama 719-0232, Japan} 
\altaffiltext{6}{Institute for the Physics and Mathematics of the 
Universe, University of Tokyo, Kashiwa, Japan}
\altaffiltext{7}{Department of Physics, Tokyo Institute of
Technology, 2-12-1 Ookayama, Meguro-ku, Tokyo 152-8551, Japan}
\altaffiltext{8}{Department of Physics, Faculty of Science,
Kagoshima University, 1-21-35 Korimoto, Kagoshima 890-0065, Japan}
\altaffiltext{9}{National Astronomical 
Observatory of Japan, Osawa, Mitaka, Tokyo 181-8588, Japan}
\altaffiltext{10}{Toyota Technical Development, Corp., 1-21 Imae,
Hanamoto-cho, Toyota, Aichi 470-0334, Japan}


\begin{abstract}
 We present early phase observations in optical and near-infrared wavelengths 
 for the extremely luminous Type Ia supernova (SN Ia) 2009dc.
 The decline rate of the light curve is $\Delta m_{15}(B)=0.65\pm 0.03$,
 which is one of the slowest among SNe Ia.
 The peak $V$-band absolute magnitude is $M_{V}=-19.90\pm 0.15$ mag
 even if the host extinction is $A_{V}=0$ mag.
 It reaches $M_{V}=-20.19\pm 0.19$ mag for the host extinction of $A_{V}=0.29$ mag 
 as inferred from the observed Na~{\sc i} D line absorption in the host. 
 Our $JHK_{s}$-band photometry shows that the SN is
 one of the most luminous SNe Ia also in near-infrared wavelengths.
 These results indicate that SN 2009dc belongs to the most luminous class of 
 SNe Ia, like SN 2003fg and SN 2006gz.
 We estimate the ejected $^{56}$Ni mass of $1.2\pm 0.3$ $\Msun$ for no
 host extinction case (or $1.6$$\pm$ $0.4$ M$_{\odot}$ for the host extinction 
 of $A_{V}=0.29$ mag).
 The C~{\sc ii} $\lambda$6580
 absorption line keeps visible until a week after maximum, which diminished 
 in SN 2006gz before its maximum brightness.
 The line velocity of Si~{\sc ii} $\lambda$6355 is about 8000 km~s$^{-1}$ 
 around the maximum, being considerably slower than that of SN 2006gz, while 
 comparable to that of SN 2003fg.
 The velocity of the C~{\sc ii} line is almost comparable to that of the
 Si~{\sc ii}. The presence of the carbon line suggests that thick 
 unburned C+O layers remain after the explosion.
 SN 2009dc is a plausible candidate of the super-Chandrasekhar mass SNe Ia.

\end{abstract}


\section{Introduction}

 Type Ia Supernovae (SNe Ia) have been believed to occur when the mass of the  
 progenitor white dwarf (WD) reaches the Chandrasekhar's limiting mass, by mass 
 accretion from a companion star.
 The homogeneity in their light curves is explained by this scenario and the
 calibrated luminosity of SNe Ia has been used as an important tool for the
 constraints on the expansion rate and the dark energy content of the universe
 (Perlmutter et al. 1999; Riess et al. 1998).
 However, their progenitors and detailed explosion mechanism
 have not been confirmed yet (e.g., Nomoto et al. 1997; Hillebrandt \& Niemeyer 2000).

 Observationally, SNe Ia have been classified into three subclasses:
 normal SNe Ia, overluminous SNe Ia (SN 1991T-like), and faint SNe Ia 
 (SN 1991bg-like) (Branch et al. 1993; Filippenko 1997; Li et al. 2001).
 The light curves of more luminous SNe Ia decline more slowly (Phillips 1993).

 Recently, two extremely luminous SNe Ia 2003fg and 2006gz have been 
 discovered (Howell et al. 2006; Hicken et al. 2007).
 Their absolute maximum magnitudes are $M_{V}=-19.94\pm0.06$ mag
 for SN 2003fg and $-19.74\pm0.16$ mag for SN 2006gz, and both 
 SNe show the slowest luminosity evolution.  
 Such an extreme brightness suggests that their progenitor's masses
 exceed the Chandrasekhar limit (``super-Chandrasekar mass WD'') 
 (Howell et al. 2006; Hicken et al. 2007).
 Interestingly, these SNe showed strong carbon absorptions 
 in their early stages; the C~{\sc ii} $\lambda6580$ line was 
 seen in SN 2003fg around maximum, while the line 
 diminished before maximum in SN 2006gz. The expansion velocity inferred from 
 the Si~{\sc ii} $\lambda$6355 line around maximum was slow in SN 2003fg
 ($\sim$ 8000 km s$^{-1}$), while it was typical ($\sim$ 11000-12000 km s$^{-1}$) 
 in SN 2006gz.

 SN 2009dc was discovered on 2009 Apr 9.31 UT at non-filter magnitude 
 of 16.5 near the outer edge of an S0 galaxy UGC 10064 
 (Puckett et al. 2009, $\mu=34.88\pm 0.15$ from the $NED$ database; 
 Falco et al. 1999).
 A follow-up observation on Apr 16.22 revealed spectroscopic similarity
 with SN 2006gz before maximum light, including the existence of
 conspicuous C~{\sc ii} features (Harutyunyan et al. 2009).
 The expansion velocity deduced from the Si~{\sc ii} $\lambda$6355
 line is about 8700 km~s $^{-1}$, 
 which is slower than that of SN 2006gz (Hicken et al. 2007), but
 comparable to that of SN 2003fg (Howell et al. 2006).

 In this Letter, we show our photometric and spectroscopic observations 
 of this peculiar SN from $-8.1$ days through $+80.5$ days after the maximum.
 The observational results strongly suggest that SN 2009dc is a 
 super-Chandrasekher SNe Ia having some peculiar properties 
 compared with other candidates.

\section{Observations and Reduction}

 We performed $BVR_{c}I_{c}$-band photometry of SN 2009dc 
 on 30 nights from 2009 
 Apr 17.8 UT ($-8.1$ days after the $B$-band maximum; see \S 3.1) 
 through Jul 14.5 ($+80.5$ days), 
 using HOWPol (Hiroshima One-shot Wide-field Polarimeter; 
 Kawabata et al. 2008) installed to the 1.5 m KANATA telescope at 
 Higashi-Hiroshima Observatory. 
 The images were reduced according to a standard procedure of 
 a CCD photometry. We performed point-spread-function fitting photometry 
 using $DAOPHOT$ package in $IRAF$.
 The magnitude is calibrated with photometric standard stars in Landolt fields  
 (Landolt 1992) observed on photometric nights.
 Additionally, we obtained $g'R_{c}I_{c}$-band photometric data 
 on 10 nights from $-1.7$ to +19.3 days, using MITSuME 0.5 m 
 telescope (Multicolor Imaging Telescopes for Survey and Monstrous Explosions).
 The MITSuME $R_{c}I_{c}$-band magnitudes are consistent with the KANATA/HOWPol 
 photometry within systematic differences less than $0.03$ mag. 

 Our near-infrared (NIR) photometric data were obtained from 
 $-2.8$ through $+44.0$ days using the 1 m telescope in Kagoshima University
 and the 1.88 m telescope at Okayama Astrophysical Observatory of NAOJ equipped 
 with ISLE (near-infrared  imager and spectrograph; Yanagisawa et al. 2006).
 Their magnitude calibrations were performed with nearby stars 
 in 2MASS catalogue.

 The optical spectra were obtained using GLOWS (Gunma LOW 
 dispersion Spectrograph) installed to the 1.5 m telescope 
 at Gunma Astronomical 
 Observatory on six nights from 
 $-3.3$ through +53.7 days. 
 The wavelength coverage was 4200--8000 \AA\ and its resolution was 
 $R=\lambda /\Delta\lambda = 330$ at 6000 \AA.
 The flux was calibrated using several spectrophotometric standard 
 stars taken in the same night.
 We have removed the strong telluric absorption features from the 
 object spectra using the standard star spectra.

 \section{Results and Discussion}

 \subsection{Light Curves}

\begin{figure}
  \begin{center}
    \begin{tabular}{c}
      \resizebox{100mm}{!}{\includegraphics{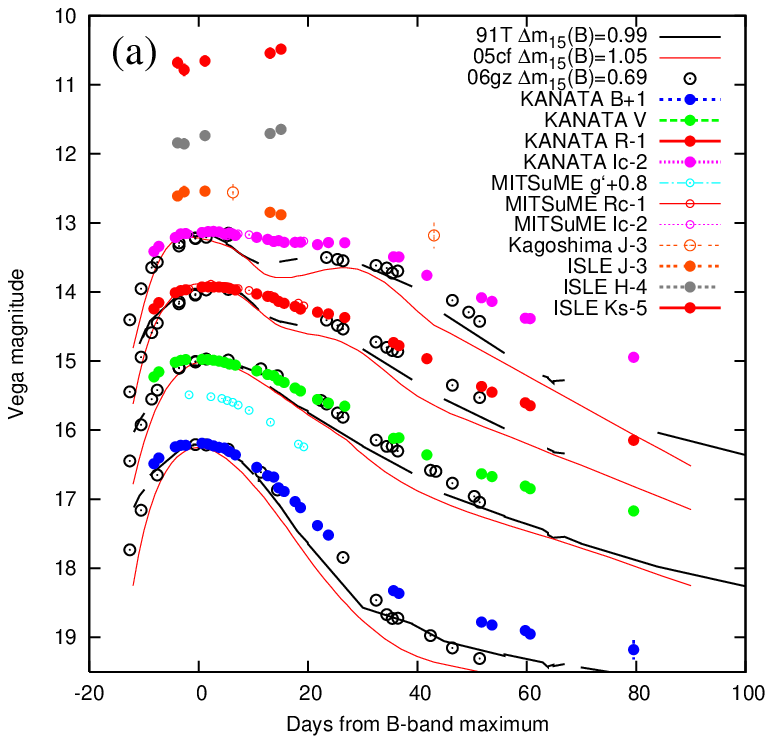}} \\
      \resizebox{100mm}{!}{\includegraphics{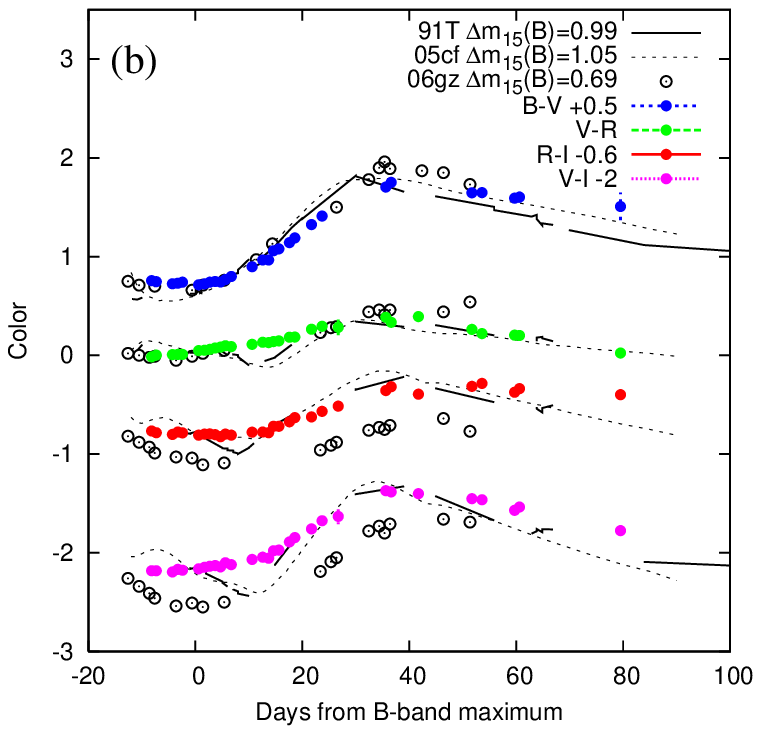}} \\
    \end{tabular}
    \caption{(a) $BVR_{c}I_{c}$-band light curves of SN 2009dc, compared with 
 a super-Chandrasekhar SN Ia 2006gz 
 (Hicken et al. 2007) and an overluminous SN Ia 
 1991T (Lira et al. 1998) and a normal SN 2005cf (Wang et al. 2009).
 The Galactic extinction has been corrected for in SN 2009dc. 
 For other SNe the light curves are shifted to match the maximum magnitude.
 (b) $B-V$, $V-Rc$, $Rc-Ic$, $V-Ic$ color evolutions, compared with
 those of SNe 1991T, 2005cf and 2006gz.  
 In SN 2009dc, only Galactic extinction ($E(B-V)=0.071$ mag and $R_{V}=3.1$) 
 is corrected for.
 In other SNe Ia, the extinctions in both our and host galaxies are corrected for
 ($E(B-V)=0.3$ mag for SN 1991T, $0.12$ for SN 2005cf, $0.12$ for SN 2006gz). 
 The color evolution of SN 2009dc is unique compared with those 
 of the other SNe Ia.}
  \end{center}
\end{figure}

  We show optical and NIR light curves of SN 2009dc in Figure 1 (a).
 The Galactic extinction of $E(B-V)=0.071$ mag and $R_{V}=3.1$
 are corrected for (see \S 3.3). We derive $B$-band maximum magnitude of 
 15.19$\pm$0.16 mag and its date of $54946.9\pm 0.2$ MJD (Apr 25.9$\pm 0.2$ UT) 
 by polynomial fitting to the light curve. 
 In Figure 1, we compare SN 2009dc with the super-Chandrasekhar 
 SN Ia 2006gz (Hicken et al. 2007; $\Delta m_{15}(B)=0.69$), 
 an over-luminous SN Ia 1991T (Lira et al. 1998; $\Delta m_{15}(B)=0.99$) 
 and a normal SN Ia 2005cf (Wang et al. 2009; $\Delta m_{15}=1.05$).

  We notice that the brightness evolution across the maximum is slower than 
 those of SNe 1991T and 2005cf in any band. 
 We derive the decline rate of $\Delta m_{15}(B)=0.65\pm 0.03$ for SN 2009dc, 
 which is similar to $\Delta m_{15}(B)=0.69\pm 0.04$ of 
 SN 2006gz (Hicken et al. 2007). The decline rate of SN 2009dc 
 is one of the smallest ones among SNe Ia which have ever been published.

  The $J$-band light curve (which would be the first NIR light curve of
 super-Chandrasekhar SNe Ia ever published) merginally shows
 a more significant dip between the first and second 
 maximum than the I-band light curve. The $H,K$-band light curves suggest the
 existence of more luminous secondary peak than the first one. These 
 characteristics are likely typical for SNe Ia 
 (Krisciunas et al. 2004; Wang et al. 2009)

\subsection{Color Evolution}

 We show the evolution of color indices of SN 2009dc in Figure 1 (b), 
 together with SNe 1991T,  2005cf and 2006gz for comparison.
 The evolution of $B-V$ of SN 2009dc is similar to those of SNe 1991T, 2005cf and 
2006gz; this suggests that the Lira-Phillips relation (homogeneous $B-V$ evolution
at 30--90 days, Phillips et al. 1999) also holds for SN 2009dc.
We discuss it in \S 3.3. On the other hand, the evolutions 
 of $V-R$, $R-I$ and $V-I$ colors in SN 2009dc are somewhat different from those of 
 SNe 1991T and 2005cf; the color indices of SN 2009dc become redder monotonously, 
 while the other SNe Ia (except for SN 2006gz) have small troughs at 10--15 
 days after the $B$-band maximum.
 SN 2006gz shows the color evolution similar to SN 2009dc, 
 while it keeps bluer at $-8$ through $+60$ days and shows broad troughs in
 the $R-I$ and $V-I$ curves.

\begin{deluxetable*}{ccccccc}
  \tabletypesize{\scriptsize}
    \tablecaption{Estimated maximum absolute magnitude, luminosity and 
     the $^{56}$Ni mass in some extinction cases \label{tbl:obs} } 
  \tablewidth{0pt}
   \startdata
    \hline
    \multicolumn{2}{c}{E$(B-V)$ (mag)} & host $A_{V}$ & $R_{V}$ & 
    $M_{V, {\rm max}}$ & $L_{{\rm max}}$ & $^{56}$Ni mass   \\
    Galactic & host & (mag) & & (mag) & (erg~s$^{-1}$) & ($\Msun$) \\ \hline \hline 
        $0.07$ & $0$       & $-$    &  $3.1$ & $-19.90\pm 0.15$  & 
        $(2.1\pm 0.5)\times 10^{43}$ & $1.2\pm 0.3$    \\
        $0.07$ &  $0.14$   & $0.29$ &  $2.1$ & $-20.19\pm 0.19$  & 
        $(2.9\pm 0.8)\times 10^{43}$ & $1.6\pm 0.4$     \\
        $0.07$ &  $0.14$   & $0.43$ &  $3.1$ & $-20.32\pm 0.19$  & 
        $(3.3\pm 0.9)\times 10^{43}$ & $1.8\pm 0.5$     
   \enddata
\tablenotetext{}{} 
\end{deluxetable*}{}

\subsection{Host Galactic Extinction and Absolute Magnitude}
 
 To confirm that SN 2009dc is one of the most luminous SNe Ia,
 it is important to determine the extinction toward this SN 
 (Galactic + host).

  The Galactic color excess is estimated to be $E(B-V)=0.071$ mag 
 (Schlegel et al. 1998), corresponding to an extinction of 
 $A_{V}=0.22$ mag within our Galaxy (a typical selective extinction 
 $R_{V}=3.1$ is assumed).
 On the other hand, the extinction within the host galaxy is 
 somewhat uncertain.
 If the Lira-Phillips relation holds for SN 2009dc, 
 it predicts a reddening of $E(B-V)=(0.37 \pm 0.08$) mag. 
  However, this is likely an overestimation because the 
 equivalent width (EW) of the Na~{\sc i} D absorption line 
 in the host galaxy ($1.0$ \AA ) is only twice the EW of Na~{\sc i} D
 in our Galaxy ($0.5$ \AA ; Tanaka et al. 2009).
 If we assume that the extinction is simply proportional to the EW, 
 $E(B-V)=0.14$ mag is plausible for the host extinction. Additionally, 
 the empirical relation between the color excess and the EW of 
 Na~{\sc i} D (Turatto et al. 2003; we adopt their
 lower extinction case) predicts $E(B-V)=0.15$ mag.
 These two values are consistent. This also suggests that the 
 relation is consistent with our Galaxy's values of $E(B-V)=0.071$ mag 
 and the EW of $=0.5$ \AA\ for the Galactic Na~{\sc i} D line.
 In Table 1 we summarize the estimated absolute magnitude for 
 various cases of extinction parameters.
 There is an additional uncertainty for the extinction 
 due to the diversity of $R_{V}$ (Wang et al. 2006; Krisciunas et al. 2006;
 Elias-Rosa et al. 2006).
 We adopt $R_{V}=2.1$ and $3.1$ following Hicken et al. (2006).
 Even if $E(B-V)=0$ mag within the host galaxy is assumed,
 the absolute maximum magnitude is $M_{V}=-19.90\pm 0.15$ mag, 
 indicating that SN 2009dc is one of the most luminous SNe Ia.

 Krisciunas et al. (2004) pointed that the absolute maximum magnitude in 
 NIR bands does not depend on the decline rate (except for faint SNe Ia) 
 and derived the mean values of $-18.6$ mag,  $-18.2$ mag and $-18.4$
 mag for $J$, $H$ and $Ks$-bands, respectively. We estimate the absolute maximum
 magnitudes of SN 2009dc as $M_{J}=-19.20\pm0.16$ mag, $M_{H}=-19.00\pm0.17$ mag,
 $M_{Ks}=-19.19\pm0.17$ mag for zero host extinction case,
 which suggests that SN 2009dc is very luminous even in NIR wavelengths.

 It is interesting to examine whether the maximum magnitude-$\Delta m_{15}(B)$ 
 relation (e.g. Altavilla et al. 2004) holds for this brightest SN Ia. 
 If there is no host extinction, the absolute V magnitude 
 derived from observations is roughly consistent ($< 1-2 \sigma$) with the relation. 
 On the other hand, it seems much brightter (by $ 3-4 \sigma$) than the prediction
 of this relation if the host extinction is $A_{V}=0.29$ mag.

\subsection{Bolometric Light Curve and $^{56}$Ni Mass}

\begin{figure}
\begin{center}
\begin{tabular}{c}
\includegraphics[scale=0.9]{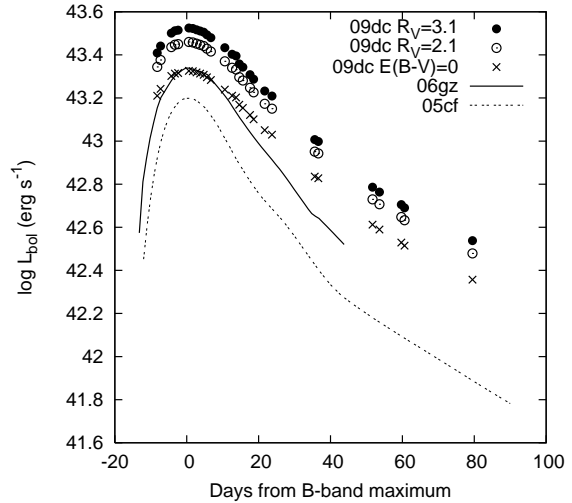}
\end{tabular}
  \caption{Bolometric light curve of SN 2009dc. 
    Filled and open circles show the cases of the host extinction 
    $E(B-V)=0.14$ mag with $R_{V}=3.1$ and $2.1$, respectively
    The asterisk shows the case with no host extinction. 
    The bolometric light curves of the normal SN Ia 2005cf 
    (dashed line; Wang et al. 2009) and the super-Chandra SN 2006gz
    (thick line; Hicken et al. 2007) are shown for comparison.} 
\end{center} 
\end{figure}

  We obtain the bolometric luminosity of SN 2009dc using our 
  $BVR_{c}I_{c}$-band data, assuming that the optical 
  luminosity occupies about 60\% of the bolometric one around 
  maximum brightness (Wang et al. 2009). 
  Because of the uncertainty involved in this assumption, 
  we consider that this bolometric luminosity may have a somewhat large 
  systematic error ($\sim 20$\%). 
  To confirm the reliability, we also calculate the bolometric luminosity 
  from $BVRI$+$JHKs$ data at $-3$ days and check the consistency. 
  We assume that the integrated 
  $BVRIJHKs$ luminosity is 80\% of the total (Wang et al. 2009). 
  They agree within an error of $\sim$ 12\%.
  But this discrepancy includes an uncertainty of determining
  the maximum date of $JHKs$-band luminosity.

  The obtained bolometric light curves are shown in Figure 2. 
  We also calculate the bolometric luminosity of SN 2005cf with the same assumption 
  and confirm the consistency with the results by Wang et al. (2009).
  Even if we assume that the host extinction is zero, 
  the maximum bolometric luminosity is
  $L_{\rm max}=(2.1\pm 0.5)\times 10^{43}$erg~s$^{-1}$, which is 
  comparable to that of SN 2006gz, $(2.18\pm 0.39)\times 10^{43}$
  erg~s$^{-1}$ for $E(B-V)=0.18$ mag (Hicken et al. 2007).
  When we adopt the host extinction of $E(B-V)=0.14$ mag and $R_{V}=3.1$,
  $L_{\rm max}=(3.3\pm 0.9)\times 10^{43}$erg~s$^{-1}$,
  which is likely to exceed that of SN 2003fg 
  ($\sim (2.5-2.8)\times10^{43}$ erg~s$^{-1}$; Howell et al. 2006).

  The mass of ejected $^{56}$Ni can be roughly estimated from 
  the peak luminosity (e.g., Arnett 1982). 
  Stritzinger \& Leibudgut (2005) suggested
  that the $^{56}$Ni mass depends approximately on 
  the peak bolometric luminosity and its rising 
 time ($t_{r}$ days), as
\begin{equation}
   L_{\rm max} = \left( 6.45~e^{\frac{-t_{r}}{8.8 {\rm d}}}+
  1.45~e^{\frac{-t_{r}}{111.3 {\rm d}}} \right)
  \left( \frac{M_{\rm Ni}}{M_{\odot}} \right) \times 10^{43} \mbox{ erg s$^{-1}$}.
\end{equation}
   The slow evolution of brightness in SN 2009dc around the maximum suggests 
  that the rising time of the bolometric luminosity is comparable to or
  slightly longer than those of SN 2006gz 
  ($\sim 18.5$ days; Hicken et al. 2007)
  or typical SNe Ia ($\sim 19$ days; e.g., Conley et al. 2006).
  Assuming $t_{r}=20$ days for SN 2009dc, 
  we derive the $^{56}$Ni mass of $1.2\pm 0.3$ $\Msun$ for no host extinction case.
  It reaches $1.8\pm 0.5$ $\Msun$ if we assume the 
  host extinction of $E(B-V)=0.14$ mag and $R_{V}=3.1$ (Table 1).
  Although the derived $L_{\rm max}$ and the $^{56}$Ni mass still 
  include somewhat large uncertainties, the observational 
  results suggest that the mass of the progenitor might exceed 
  the Chandrasekhar-limit.

 \subsection{Spectral Evolution}

 \begin{figure}
 \begin{center}
 \begin{tabular}{c}
 \includegraphics[scale=0.8]{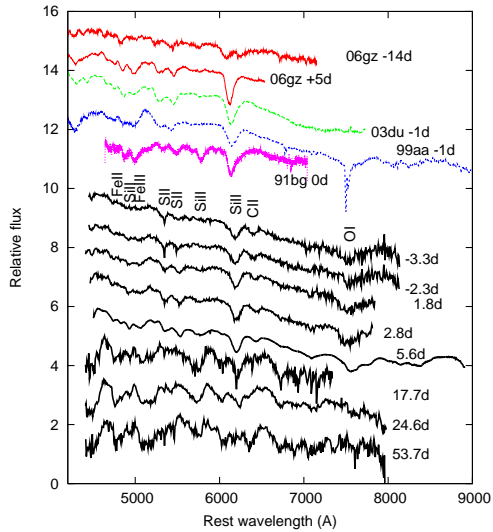}
 \end{tabular}
 \caption{Spectra of SN 2009dc compared with those of
 other SNe Ia; subluminous object SN 1991bg, 
 over-luminous object SN 1999aa, 
 typical object SN 2003du and the super-Chandrasekhar
 SN Ia 2006gz. All spectra are calibrated to the Rest frame wavelength.
 The spectrum of SN 2009dc at $+5.5$ days is from Tanaka et al.
 (2009). The telluric absorption feature have been removed.}

 \end{center}
 \end{figure}

 \begin{figure}
 \begin{center}
 \begin{tabular}{c}
 \includegraphics[scale=1.0]{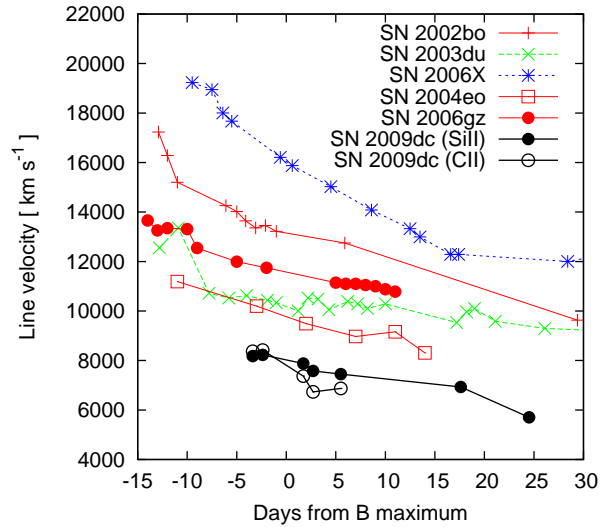}
 \end{tabular}
 \caption{Si~{\sc ii} $\lambda$6355 line velocity evolution of SN 2009dc and
 comparative SNe Ia, 2006gz (Hicken et al. 2007), 2006X 
 (Yamanaka et al. 2009), 2004eo (Pastorello et al.2007), 
 2003du (Stanishev et al.2007),  2002bo (Benetti et al.2004).
 We also show the C~{\sc ii} $\lambda$6580 line velocity of SN 2009dc
 with black open circles. The low expansion velocity of SN 2009dc is remarkable.} 
 \end{center} 
 \end{figure}

 In Figure 3, we compare the spectra of SN 2009dc with those of the 
 super-Chandrasekhar SN 2006gz at $-14$ days and $+5$ days (Hicken et al. 2007), 
 the subluminous  SN 1991bg (Filippenko et al.1992), the overluminous SN 1999aa 
 (Garavini et al.2004) and the typical SN 2003du (Stanishev et al.2007) 
 around maximum.
 The spectra of SN 2009dc around maximum show  
 Si~{\sc ii} $\lambda$6355 absorption, a W-shape S~{\sc ii} absorption
 feature and Fe-group multiple absorptions.
 Additionally, the absorption line of C~{\sc ii} $\lambda$6580 is
 seen in SN 2009dc, even a few days after the maximum.
 This feature is seen only in a small fraction of SNe Ia at their
 earliest epochs (Tanaka et al. 2008).
 In the super-Chandrasekhar candidate SN 2006gz, the carbon feature 
 also exists, while the feature is significant only in the
 earliest stages ($\lesssim -10$ days; Hicken et al. 2007).
 In another, more distant super-Chandrasekhar SN 2003fg,
 the C~{\sc ii} $\lambda$6580 feature is not significant 
 at $+2$ days, while there is 
 a possible carbon feature around 4150 \AA\ at the same epoch. 
 These indicate that a massive C+O layer exists
 in the atmosphere of SN 2009dc. 
 
 In Figure 4, we show the line velocity of Si~{\sc ii} $\lambda$6355
 together with those in other SNe Ia.
 The Si~{\sc ii} line velocity of SN 2009dc is $\sim$ 8000 km~s$^{-1}$
 at $-4$ days and then decreases to $\sim$ 6000 km~s$^{-1}$ by $+24$ days.
 This indicates that SN 2009dc is one of the most slowly expanding SNe Ia
 (except for faint SNe Ia). The line velocity is much lower than that of SN 2006gz, 
 while comparable with that of SN 2003fg ($8000\pm 500$ km s$^{-1}$ 
 at $+2$ days; Howell et al. 2006).
 The velocity of the C~{\sc ii} $\lambda$6580 line in SN 2009dc evolves
 roughly as well as that of the Si~{\sc ii} line (Fig. 4).

\section{Discussion and Conclusions}

 We summarize the observational characteristics of the peculiar 
 SN Ia 2009dc as follows:
 (1) one of the slowest evolution of the light curve, i.e.,  
 $\Delta m_{15}(B)=0.65\pm 0.03$, 
 (2) one of the most luminous SNe Ia, i.e., $M_{V}=-19.90\pm 0.15$ or brighter,
 (3) a strong carbon feature in the early spectra, and
 (4) the lowest expansion velocity among normal and overluminous SNe Ia.
 The first three features are similar to another super-Chandrasekhar 
 candidate SN 2006gz, while the last item is clearly different.
 Although the detailed data for the distant super-Chandrasekhar candidate
 SN 2003fg are lacking, its expansion velocity is comparable to 
 that of SN 2009dc.
 The C~{\sc ii} $\lambda$6580 feature is still present at $+5$ days
 in SN 2009dc, although it diminishes by the similar epoch in SNe 2003fg 
 and 2006gz.
 These facts suggests an existence of a massive unburned C+O layer
 in the ejecta of SN 2009dc.
 Additionally, we derive the amount of ejected $^{56}$Ni mass of 
 $1.2\pm 0.3$ $\Msun$ even for no host extinction.
 If we assume the host extinction of $E(B-V)=0.14$ mag and $R_{V}=2.1$,
 it reaches $1.6\pm 0.4$ $\Msun$.
 Therefore, we suggest that SN 2009dc is a SN Ia explosion with
 a super-Chandrasekhar mass WD.

 \acknowledgements
 This research has been supported in part by the Grant-in-Aid for 
 Scientific Research from JSPS (20540226, 20740107, 21018007,
 20840007) and MEXT (19047004, 20040004), and WPI Initiative, MEXT.
 M.T. has been supported by the JSPS Research Fellowship for Young
 Scientists.

\end{document}